\documentclass[12pt,thmsa]{article}

\usepackage{cite}
\usepackage{amsmath}

\begin{document}

\title{On a simple approach to nonlinear oscillators}
\author{Francisco M. Fern\'{a}ndez \\
INIFTA (UNLP, CCT La Plata--CONICET), Divisi\'{o}n Qu\'{i}mica Te\'{o}rica,\\
Diag. 113 y 64 (S/N), Sucursal 4, Casilla de Correo 16,\\
1900 La Plata, Argentina \\
e--mail: fernande@quimica.unlp.edu.ar}
\date{}
\maketitle

\begin{abstract}
We analyze a simple textbook approach to nonlinear oscillators
proposed recently, disclose its errors, limitations and
misconceptions and complete the calculations that the authors
failed to perform.
\end{abstract}

\section{Introduction}

\label{sec:intro}

In a recent article Ren and He\cite{RH09} proposed a simple method
for the approximate calculation of the period of nonlinear
oscillators and applied it to three rather trivial toy models. In
this paper we discuss this remarkable scientific contribution that
is another illustrative example of the new trend in mathematical
physics promoted by certain journals.

\section{Their method and our improvements}

\label{sec:method}

Ren and He\cite{RH09} chose the dimensionless equation of motion
\begin{equation}
u^{\prime \prime }+f(u)=0  \label{eq:eq_motion}
\end{equation}
with the initial conditions
\begin{equation}
u(0)=A,\,u^{\prime }(0)=0  \label{eq:ini_cond}
\end{equation}
where the prime indicates differentiation with respect to the independent
variable $t$.

The authors gave the following recipe:\ ``We always choose cosine or sine
function as a trial function for nonlinear oscillators. For the above
initial conditions, we choose
\begin{equation}
u=A\cos \omega t  \label{eq:u[1]}
\end{equation}
where $\omega $ is the angular frequency of the nonlinear oscillator to be
further determined. Substituting Eq.~(\ref{eq:u[1]}) into Eq.~(\ref
{eq:eq_motion}) results in''
\begin{equation}
u^{\prime \prime }+f(A\cos \omega t)=0  \label{eq:mot_wrong}
\end{equation}
It is unbelievable that the authors did not realize that this equation does
not apply to an arbitrary nonlinear oscillator for all values of $t$. In
fact it is suitable only for the Hooke's force $f(u)=\omega ^{2}u$. The
authors proceeded as follows: ``Integrating Eq.~(\ref{eq:mot_wrong}) twice
with respect to $t$, we have
\begin{equation}
u^{\prime }=-\int_{0}^{t}f(u)\,dt  \label{eq:u'_He}
\end{equation}
and''
\begin{equation}
u(t)=\int_{0}^{t}u^{\prime }\,dt  \label{eq:u_wrong}
\end{equation}
Another incredible mistake, this equation does not satisfy the first initial
condition in Eq~(\ref{eq:ini_cond}).

In the discussion and conclusions Ren and He\cite{RH09} admirably argued
that ``For an oscillator with initial conditions (\ref{eq:ini_cond}) we
have''
\begin{equation}
u(0)=A,\,u(T/4)=0,\,u(T/2)=-A,\,u(3T/4)=0,\,u(T)=A  \label{eq:further_cond}
\end{equation}
However, they did not bother to make it explicit that this set of
equations is valid only for a nonlinear oscillator with odd force
$f(-u)=-f(u)$. It is also probable that they were not aware of
this obvious fact. Fortunately the authors only treated such
particular cases.

Before proceeding with the discussion of this admirable piece of work, we
recall a well known result in classical dynamics. If we multiply Eq.~(\ref
{eq:eq_motion}) by $u^{\prime }$ and integrate we obtain the textbook
expression
\begin{equation}
\frac{u^{\prime 2}}{2}+V(u)=E  \label{eq:E}
\end{equation}
where $E$ is a constant of the motion and $dV/du=f$.

To continue with the discussion of the paper by Ren and He\cite{RH09} we
derive the correct expression that in our opinion is the basis of their
approach. If we integrate Eq.~(\ref{eq:eq_motion}) twice and take into
account the boundary conditions already indicated above we obtain
\begin{equation}
u(t)=A-\int_{0}^{t}\int_{0}^{t^{\prime }}f(u(t^{\prime \prime
}))\,dt^{\prime \prime }\,dt^{\prime }  \label{eq:u(t)}
\end{equation}
that those authors supposedly tried to derive. The main idea
behind their approach is that if one introduces an approximate
solution into the right--hand side of Eq.~(\ref{eq:u(t)}) the
result is expected to be an improvement. To obtain an approximate
analytical expression for the period $T $ the authors resorted to
the condition $u(T/4)=0$ and in the conclusions they stated that
``The suggested solution procedure is valid for conservation
systems with unchanged amplitude''. In fact, any equation of the
form (\ref{eq:eq_motion}) represents a conservative system and
exhibits a constant amplitude given by $V(u)=E$ that determines
the turning points. Did the authors know it?. As indicated above
$u(T/4)=0$ is valid only for symmetric problems $V(-u)=V(u)$ or
odd functions $f(-u)=-f(u)$. In fact, they only considered
examples with that property without stating it explicitly in the
general presentation of the approach. To facilitate the discussion
below it is worth noticing that for an odd force $A\geq u(t)\geq
0$ if $0\leq t\leq T/4$.

\section{Their examples and our improvements}

\label{sec:examples}

The first example chosen by Ren and He\cite{RH09} is the well known and
widely studied Duffing oscillator:
\begin{equation}
f(u)=u+\epsilon u^{3}  \label{eq:Duffing}
\end{equation}
Undergraduate students learn how to obtain satisfactory
approximate solutions to this equation in most textbooks on
classical mechanics. As shown below, the results of Ren and
He\cite{RH09} are of such kind, except that they were published in
a research journal and were derived in a sloppy way. Apparently,
those authors did not realize that the model parameters $\epsilon
$ and $A$ do not appear separately in the results but in the form
of the only relevant quantity $\rho =\epsilon A^{2}$, an
observation that greatly facilitates the discussion of the
results. They first substituted
Eq.~(\ref{eq:u[1]}) with $\omega =2\pi /T$ into the right--hand side of Eq.~(%
\ref{eq:u(t)}) and obtained an improved trajectory as well as the
approximate period\cite{RH09}
\begin{equation}
T^{[1]}(\rho )=\frac{2\pi }{\sqrt{1+\frac{7}{9}\rho }}  \label{eq:T[1]}
\end{equation}
This expression provides a reasonable approach to the actual period for all
values of $\rho $, and in particular an acceptable estimate of the limit
\begin{equation}
\lim_{\rho \rightarrow \infty }\sqrt{\rho }T(\rho )=T_{\infty }=7.416298709
\label{eq:T_inf}
\end{equation}
In fact, Eq.~(\ref{eq:T[1]}) gives us $T_{\infty }^{[1]}\approx
7.12$.

How did they derive a reasonable result from the wrong equation
(\ref {eq:u_wrong})? Simply by the addition of another error that
corrects the first one. Notice that the left-- and right--hand
sides of equations (10) and (11) in their paper do not match. They
added convenient integration constants in the last steps to
satisfy the boundary conditions.

In order to improve this first estimate Ren and He\cite{RH09} then tried the
ansatz
\begin{equation}
u(t)=A_{1}\cos (\omega t)+A_{2}\cos (3\omega t)  \label{eq:u[2]}
\end{equation}
where $A_{1}+A_{2}=A$. The additional coefficient requires an additional
condition and the authors chose
\begin{equation}
u^{\prime \prime }(0)+f(A)=0  \label{eq:second_condition}
\end{equation}
In this way they derived a complicated system of equations that suspiciously
they did not try any further. Besides, at first glance their results do not
appear to be functions of $\rho $ alone. For this reason in what follows we
derive a suitable expression and verify if it is more accurate than the
first approximation discussed above. Starting from the same premises our
expression for the period results to be
\begin{eqnarray}
&&125T^{8}\rho \left( \rho +1\right) ^{3}-7656\pi ^{2}T^{6}\rho \left( \rho
+1\right) ^{2}+120\pi ^{4}T^{4}\left( \rho +1\right) \left( 1607\rho
+735\right)  \nonumber \\
&&-64\pi ^{6}T^{2}\left( 48851\rho +55125\right) +12700800\pi ^{8}=0
\label{eq:T[2]}
\end{eqnarray}
This equation has four real roots $-T^{[2,2]}(\rho
)<-T^{[2,1]}(\rho )<0<T^{[2,1]}(\rho )<T^{[2,2]}$, and
$T^{[2,1]}(\rho )$ is the desired approximation that satisfies
$T^{[2,1]}(0)=2\pi $.

A straightforward calculation shows that the approximation to $T_{\infty }$
is a root of

\begin{equation}
125T_{\infty }^{8}-7656\pi ^{2}T_{\infty }^{6}+192840\pi ^{4}T_{\infty
}^{4}-3126464\pi ^{6}T_{\infty }^{2}+12700800\pi ^{8}=0  \label{eq:T[2]_inf}
\end{equation}
We thus obtain $T_{\infty }^{[2,1]}\approx 7.44$ that is in fact
more accurate than $T_{\infty }^{[1]}$.

If the Duffing equation is a textbook exercise, the second example
studied by Ren and He\cite{RH09}, given by
\begin{equation}
f(u)=\epsilon \ \mathit{sgn}(u)  \label{eq:ex_2}
\end{equation}
is ridiculously trivial because it is almost impossible to miss the exact
result (remember that this paper is published in a research journal). We
know that $\mathit{sgn}(u)=1$ for $0\leq t\leq T/4$; therefore,  if we
substitute $f(u)=\epsilon $ into the right hand--side of Eq.~(\ref{eq:u(t)})
we obtain $u(t)=A-\epsilon t^{2}/2$ and the exact period $T=4\sqrt{%
2A/\epsilon }$. Any trial function that is positive definite in this
interval leads to the same result, even if it does not satisfy the initial
condition.

The third example is
\begin{equation}
f(u)=\omega _{0}^{2}u+\epsilon u|u|  \label{eq:ex_3}
\end{equation}
but it is sufficient to consider $f(u)=\omega _{0}^{2}u+\epsilon u^{2}$ by
virtue of the argument given above. Notice that in this case the independent
model parameters are $\omega _{0}$ and $\rho =\epsilon A$. If we choose the
trial function (\ref{eq:u[1]}) we obtain their result
\begin{equation}
T^{[1]}=\frac{2\pi }{\sqrt{\omega _{0}^{2}+\frac{4+\pi ^{2}}{16}\rho }}
\label{eq:T[1]_3}
\end{equation}
from which it follows that $T_{\infty }^{[1]}\approx 6.75$.

If we use the trial function (\ref{eq:u[2]}) we easily obtain and improved
expression for the period
\begin{eqnarray}
&&T^{6}\rho \left( \omega _{0}^{4}+2\omega _{0}^{2}\rho +\rho ^{2}\right)
\left( 9\pi ^{2}-16\right)   \nonumber \\
&&-8\pi ^{2}T^{4}\left[ 256\omega _{0}^{4}+15\omega _{0}^{2}\rho \left( 3\pi
^{2}+16\right) +\rho ^{2}\left( 45\pi ^{2}-16\right) \right]   \nonumber \\
&&+16\pi ^{4}T^{2}\left[ 5120\omega _{0}^{2}+\rho \left( 369\pi
^{2}+1136\right) \right] -294912\pi ^{6}=0  \label{eq:T[2]_3}
\end{eqnarray}
and the corresponding limit $\rho \rightarrow \infty $%
\begin{equation}
T_{\infty }^{6}\left( 9\pi ^{2}-16\right) +8\pi ^{2}T_{\infty }^{4}\left(
16-45\pi ^{2}\right) +16\pi ^{4}T_{\infty }^{2}\left( 369\pi
^{2}+1136\right) -294912\pi ^{6}=0  \label{eq:T[2]_inf_3}
\end{equation}
In this case there are only two real roots $-T_{\infty }^{[2]}$ and $%
T_{\infty }^{[2]}\approx 6.867$ and the agreement of the positive
one with the exact result $T_{\infty }=6.868663935$ is remarkable.

\section{Conclusions}

The method proposed by Ren and He\cite{RH09} could be thought of
as an undergraduate exercise on classical mechanics with the
limitation
that it is valid only for odd forces because of the necessary condition $%
u(T/4)=0$. Their paper may be of some pedagogical value as an exercise for
undergraduate students who may be asked to find as many mistakes as
possible. The lecturer may even organize a kind of competition among groups
of students dedicated to such a task. We think that it can be a hilarious
class. The second example in that paper is trivial and, therefore, only
useful as a curiosity  for beginners.

We have carried out the second--order approximation for the Duffing
oscillator that those authors left unfinished, and also showed how to do
that calculation for the third example that they never tried. There is no
need to say that the results of that paper have no serious utility
whatsoever for actual research in the field of nonlinear oscillations.
However, it is not surprising that such a sloppy paper had been published in
a research journal where one finds many such examples. We have in fact
discussed several of them in a series of communications\cite
{F07,F08b,F08c,F08d,F08e,F08f,F09a,F09b,F09c}. In particular we want to draw
the reader's attention to the extraordinary case of a predator--prey model
that predicts a negative number of rabbits\cite{F08d}.

Finally, we mention that when a mild version of this article was
submitted to the journal we were told `` We urge you to contact
the authors of this article before you submit a comment for
publication. This approach is probably going to be much more
productive for all concerned. I am going to reject the manuscript
for now, but if after speaking directly with the authors you've
come to an agreement that this manuscript should be published then
you can resubmit.'' If we understand it clearly we are urged to
ask permission from the authors to criticize their paper. Who
would agree to it?

\end{document}